\documentclass{iopart}

\usepackage[english]{babel}
\usepackage{graphicx}
\usepackage{psfrag}
\usepackage{amssymb}
\usepackage{amscd}
\usepackage{eucal}
\usepackage{color}
\usepackage{bm}
\usepackage{lipsum}
\usepackage{hyperref}
\usepackage{todonotes}
\usepackage{iopams}

\begin{document}

\title[System Size Identification from Sinusoidal Probing]{System Size Identification from Sinusoidal Probing in Diffusive Complex Networks}
\author{Melvyn Tyloo\textsuperscript{1} and Robin Delabays\textsuperscript{2,3}}
\address{\textsuperscript{1} School of Engineering, University of Applied Sciences of Western Switzerland HES-SO, CH-1951 Sion, Switzerland.}
\address{\textsuperscript{2} Automatic Control Laboratory, ETH Zurich, CH-8092 Zurich, Switzerland}
\address{\textsuperscript{3} Center for Control, Dynamical Systems and Computation, UC Santa Barbara, Santa Barbara, CA 93106-5070, USA.}
\ead{melvyn.tyloo@hevs.ch, robindelabays@ucsb.edu}

\begin{abstract}
 One of the most fundamental characteristic of a complex system is its size (or volume), which, in many modelling, is represented by the number of its individual components. 
 Complex systems under investigation nowadays are typically large and/or time-varying, rendering their identification challenging. 
 We propose here an accurate and efficient method to determine the size of (i.e., number of agents in) a complex, diffusively coupled dynamical system, that leverages the response of the system to an injected probing signal. 
 For our method to be applicable, we rely on some assumptions on system's characteristics, namely, on the spectrum of the coupling graph and on the basin stability of its steady state(s). 
 Even though such assumptions imply that our method cannot be applied to any instance of diffusively coupled group of dynamical agents, we argue that it covers relevant and interesting examples. 
 Furthermore, the simplicity of the approach and its low computational complexity renders it very interesting for the systems to which it applies. 
\end{abstract}


\section{Introduction} 
At the era of \emph{big data}, access to high-resolution (in space and time) measurements is increasingly easy. 
In parallel, the ever-improving computational power of computers allows for the analysis of such large sets of data~\cite{Hil11}. 
Among many examples, these considerations apply to domains ranging from electrical grid to social networks or gene regulatory networks. 
With \emph{maching learning} leading the way, modern technologies promise to improve the wellbeing of humanity~\cite{For16}, e.g., by leveraging the increasing amout of data gathered by \emph{Phasor Measurements Units}~\cite{Mac08} or online social networks~\cite{Sek16}, to name but a few.

The other side of the coin of this tremendous amount of data is that it comes with significant uncertainties. 
Indeed, in general, it is not possible to guarantee that each component of a large system will be monitored at all time, either due to the number of units to be monitored or to the time-varying nature of the system's components. 
These unavoidable inaccuracies and uncertainties can jeopardize the efficacy of data-based technologies. 

In this scope, it is of particular interest to be able to recover a system's characteristics (parameters, internal structure, etc.) from measured data. 
Improving the efficiency of such processes tends to increase the resolution of the inferred characteristics, both in space and time. 
For instance, recovering the underlying structure of a network of dynamical agents, based on measurements, has been an active topic of research along the last decades~\cite{Prz16,Yu06,Tim07,New18,Pei19,Tim14,Bru18,Tyl20b}. 
However, the majority of those approaches rely on the knowledge of most (if not all) the agents composing the system. 
Sometimes, a subset of these agents is not accessible to the observer, rendering the recovery of the network harder, if not impossible~\cite{Su12}. 
In the worst situations, it even happens that the oberver does not even know the actual size of the system~\cite{Hae19}. 

The number of components in a system is one of its most fundamental characteristics and is often unknown, especially for large, time-varying systems. 
A crucial step in the process of system identification is then to recover this number as accurately and with as few measurements as possible. 
Despite its apparent simplicity, this problem is actually not trivial and surpizingly underinvestigated given its fundamental relevance in the scope of system identification. 
The authors of~\cite{Su12} proposed an approach to locate hidden nodes in a networked dynamical system, which relies on the comparison between measured and predicted trajectories of the accessible agents of the system. 
This method requires a good knowledge of the differential equations determining the dynamics of the system in order to numerically integrate them. 
Moreover, the authors assume that the number of unaccessible agents is very small (usually only one). 

As far as we can tell, the most up-to-date approach to recover the number of agents in a system has been detailed in~\cite{Hae19}. 
This approach recovers the number of units as the rank of a \emph{detection matrix} constructed with time series of the measured units. 
Namely, it requires to observe the system at $k$ time steps, along $M$ different trajectories, and if $k$ and $M$ are large enough, in principle, the total number of dynamical units can be recovered. 
In summary, if $N$ is the number of observable units, this method relies on $NkM$ measured values and requires to be able to set the system in $M$ different initial conditions. 
A recent Letter~\cite{Por20} elegantly draws the link between the detection matrix of~\cite{Hae19} and the \emph{observability matrix} commonly used in control theory~\cite{Rug96}. 
Ref.~\cite{Por20} shows that the approach proposed in~\cite{Hae19} can unambiguously determine the system size if and only if the system is \emph{observable} in the control-theoretic sense. 

There is a wide range of physical or virtual systems that are typically modeled as a set of dynamical agents interacting through a diffusive coupling and according to an interaction network. 
Examples thereof range from electrical power grids~\cite{Mac08,Dor13} to opinion dynamics~\cite{Sek16,Bau20}, through models of gene regulatory networks~\cite{Bra03,Kur84b,Zha09}, control of vehicular platoons~\cite{Lin12,Gru18}, or flows in transport networks~\cite{Cor10,Kat10}.
We show here that, for such diffusively coupled systems, an accurate determination of the number of units can be performed efficiently. 
Namely, we proceed via measurement of the trajectory of one dynamical unit, following the injection of a probing signal at a single source, which reduces significantly the number of measurements compared to the state-of-the-art literature~\cite{Hae19}. 

The cost of our approach is that we require to be able to inject a probing signal at one of the nodes of the network and that it is restricted to diffusive couplings between agents. 
The probing signal needs to be tailored so as to satisfy a series of assumptions that guarantee our method to be effective. 
Namely, the amplitude of the probing should be large enough to be detected, but small enough not to destabilize the system. 
Also, the probing's frequency should be small enough, so it impacts the whole network. 
This series of assumption might seem restrictive. 
Nevertheless, we argue that our method still covers a significant range of relevant dynamical systems, power grids being just one example of those. 
Furthermore, in our opinion, the limitations above are counterbalanced by a significant reduction in computational cost and extended scalability, as only one time series has to be analyzed, and the method is independent of the system's size.

\section{Systems of diffusively coupled units}
Let us consider a system of $n$ dynamical units, interacting through a (possibly nonlinear) diffusive coupling, 
\begin{eqnarray}\label{eq:dyn}
 \dot{x}_i = \omega_i - \sum_{j=1}^n a_{ij}f_{ij}(x_i-x_j) + b_i\, , \qquad i = 1,...,n\, ,
\end{eqnarray}
where $x_i\in \mathcal{M}$ is the time-varying value of the $i$th agent, evolving on a one-dimensional manifold $\mathcal{M}$, $\omega_i\in\mathbb{R}$ is the natural, constant driving term of agent $i$, and $b_i$ will be used as an input to the system. 
Two agents $i$ and $j$ are interacting if a link between them exists in the interaction network, i.e., if and only if the corresponding term of the adjacency matrix $a_{ij}=1$. 
We assume that the interaction graph is connected if undirected and strongly connected if directed [otherwise we restrict ourselves to a (strongly) connected component].
The interaction function between $i$ and $j$ is a differentiable function $f_{ij}\colon\mathbb{R}\to\mathbb{R}$ satisfying $f_{ij}(0)=0$ and $f_{ij}'(0)>0$ (where the prime denotes the first derivative). 

If a fixed point $\bm{x}^*\in \mathcal{M}^n$ exists, one can linearize Eq.~(\ref{eq:dyn}) around it, which yields, for a small deviation $\bm{\delta}=\bm{x}-\bm{x}^*$, to the approximate dynamics
\begin{eqnarray}\label{eq:dyn_lin}
 \dot{\bm{ \delta}} = -\mathbb{J}(\bm{x}^*)\bm{\delta} + \bm{b}\, ,
\end{eqnarray}
where we define the Jacobian matrix of Eq.~(\ref{eq:dyn}), 
\begin{eqnarray}\label{eq:jac}
 \mathbb{J}_{ij}(\bm{x}^*) = \left\{
 \begin{array}{ll}
  -a_{ij}f_{ij}'(x_i^*-x_j^*)\, , & \textrm{if } i\neq j\, , \\
  \displaystyle \sum_{k\neq i} f_{ik}'(x_i^*-x_k^*)\, , & \textrm{if } i=j\, .
 \end{array}
 \right.
\end{eqnarray}
Eq.~(\ref{eq:dyn_lin}) is a good approximation of Eq.~(\ref{eq:dyn}) as long as the system remains in the vicinity of $\bm{x}^*$. 
One can verify that the structure of the interactions implies that the Jacobian $\mathbb{J}$ is a weighted directed Laplacian matrix of the interaction graph. 
Let us denotes its right- (resp. left-) eigenvectors $\bm{u}_1,...,\bm{u}_n$ (resp. $\bm{v}_1,...,\bm{v}_n$). 
From now on, we will focus on stable fixed points of Eq.~(\ref{eq:dyn}), implying that the eigenvalues have nonnegative real part, $0=\lambda_1<{\rm Re}(\lambda_2)\leq ...\leq {\rm Re}(\lambda_n)$ (note that one eigenvalue is always zero) and the eigenvectors form a basis of $\mathbb{R}^n$.

Equation~(\ref{eq:dyn_lin}) is then solved by expanding the deviation $\bm{\delta}$ over the right-eigenvectors $\bm{u}_\alpha$ of $\mathbb{J}$, i.e., $\delta_i(t) = \sum_\alpha c_\alpha(t)u_{\alpha,i}$, yielding a set of ordinary differential equations in $c_\alpha$, 
\begin{eqnarray}
 \dot{c}_\alpha(t) = -\lambda_\alpha c_\alpha(t) + \bm{v}_\alpha\bm{b}(t)\, ,
\end{eqnarray}
where we used the bi-orthogonality relation $\bm{v}_\alpha\bm{u}_\beta = \delta_{\alpha \beta}$. 
The latters are solved by multiplying by $e^{\lambda_\alpha t}$ and then integrating by parts. 
It follows that 
\begin{eqnarray}\label{eq:sols}
 c_\alpha(t) = e^{-\lambda_\alpha t}\int_0^te^{\lambda_\alpha s}\bm{v}_\alpha \bm{b}(s){\rm d}s\, , \qquad \alpha=1,...,n.
\end{eqnarray}

\section{Sinusoidal probing and system size estimation}
Our method relies on the measurement of the response of the system after the injection of a probing signal at some point. 
Such approach are used, for instance, to identify some eigenmodes in electrical grids~\cite{Pie10}, or in gene regulatory networks to infer the existence of causal relations between genes~\cite{Wag01}.
Before probing the system, we require it to be at (or close to) steady state, and if necessary, we let the transient due to initial conditions decay. 
If the system does not converge to a steady state, our method cannot be applied, but in many cases, determining the number of agent is a very secondary problem if the system is not at steady state, e.g., power grids.

\subsection{System size estimation}
In theory, any signal could be decomposed as sum of sinusoidal signals, by Fourier Transform. 
We then propose to inject the most fundamental building block of the Fourier decomposition, i.e., a sinusoidal signal at agent $i$ and to measure its impact at agent $j$ (possibly equal to $i$). 
Let 
\begin{eqnarray}\label{eq:probing}
 b_i(t) = b_0\sin(\omega_0t)\, ,
\end{eqnarray}
be the probing signal at agent $i$, and $b_j\equiv 0$ for $j\neq i$. 
The choice of amplitude $b_0$ and frequency $\omega_0$ are mostly constrained by the nature of the system under investigation. 
The probing amplitude should be small enough in order not to jeopardize the normal operation of the system, but, at the same time larger than the ambient noise to which the sytem is subjected. 
The frequency of the forcing should be sufficiently small, such that the disturbance propagates to the whole system. 
The probing is designed to leave the system in the vicinity of the fixed point $\bm{x}^*$, so that the response of the system is dominated by the linear terms and the method can be applied to general diffusive couplings. 
We defer a thorough discussion about the tayloring of the probing signal to Sec.~\ref{ssec:probing_design}. 

Introducing Eq.~(\ref{eq:probing}) into Eq.~(\ref{eq:sols}), and recombining the eigenmodes yields the following response measured at agent $j$ while probing at agent $i$, 
\begin{eqnarray}
 \delta_j^i(t) &= \sum_\alpha v_{\alpha,i}u_{\alpha,j} b_0 e^{-\lambda_\alpha t} \int_0^t e^{\lambda_\alpha s}\sin(\omega_0 s){\rm d}s \nonumber\\
 &= \sum_\alpha \frac{v_{\alpha,i}u_{\alpha,j} b_0}{\lambda_\alpha^2+\omega_0^2}  \left[\lambda_\alpha\sin(\omega_0 t) + \omega_0 e^{-\lambda_\alpha t}- \omega_0 \cos(\omega_0 t) \right] \, . \label{eq:trajs_brut}
\end{eqnarray}
Provided the probing frequency is small compared to the Jacobian's eigenvalues, $\omega_0\ll{\rm Re}(\lambda_\alpha)$ [in practice, $\omega_0<{\rm Re}(\lambda_\alpha)$], after a few periods of the probing, the exponential in Eq.~(\ref{eq:trajs_brut}) vanishes [${\rm Re}(\lambda_\alpha)t\gg 1$], and the sine dominates all non-zero modes. 
One then approximates Eq.~(\ref{eq:trajs_brut}) as
\begin{eqnarray}
 \delta_j^i(t) &= \frac{b_0}{n\omega_0}\left[1 - \cos(\omega_0t)\right] \nonumber\\ &+ \sum_{\alpha\geq 2} \frac{v_{\alpha,i}u_{\alpha,j} b_0}{\lambda_\alpha^2+\omega_0^2}  \left[\lambda_\alpha\sin(\omega_0 t) + \omega_0 e^{-\lambda_\alpha t}- \omega_0 \cos(\omega_0 t) \right] \label{eq:trajs} \\
 &\approx \frac{b_0}{n\omega_0}\left[1 - \cos(\omega_0 t)\right] + b_0\sin(\omega_0 t)\sum_{\alpha\geq 2}\frac{v_{\alpha,i}u_{\alpha,j}}{\lambda_\alpha} \\ 
 &= \frac{b_0}{n\omega_0}[1-\cos{(\omega_0 t)]} + \mathbb{J}_{ij}^{\dagger}b_0 \sin{(\omega_0 t)}\, , \label{eq:trajs2}
\end{eqnarray}
where the $\dagger$ denotes the Moore-Penrose pseudo-inverse. 
Defining
\begin{eqnarray}
 C_{ij} &= \frac{b_0}{n\omega_0}\sqrt{(\mathbb{J}_{ij}^\dagger n\omega_0)^2 + 1}\, , \quad \phi_{ij} &= \arctan\left[(\mathbb{J}_{ij}^\dagger n\omega_0)^{-1}\right]\, ,
\end{eqnarray}
we re-write Eq.~(\ref{eq:trajs2}) as
\begin{eqnarray}\label{eq:trajs3}
 \delta_j^i(t) &= \frac{b_0}{n\omega_0} + C_{ij}\sin(\omega_0 t-\phi_{ij})\, .
\end{eqnarray}

We notice that the value of the response $\delta_j^i$ averaged over an integer number of periods is $\hat{\delta}=b_0/n\omega_0$. 
Keeping track of one trajectory, we can then estimate the number of agents composing the system as 
\begin{eqnarray}\label{eq:node_est}
 \hat{n} = b_0/\omega_0\hat{\delta}\, ,
\end{eqnarray}
and the longer the measurements, the better the estimate. 

{\bf Remark.}
{\it For $\omega_0$ sufficiently small, $C_{ij}$ is also well approximated by $b_0/n\omega_0$, and thus the amplitude of the trajectory of $\delta_j^i$ is given by $2\hat{\delta}$, which is another way to estimate $n$, }
\begin{eqnarray}\label{eq:node_est_alt}
 \hat{n} &= \frac{2b_0}{\omega_0\left(\max_t\delta_j^i - \min_t\delta_j^i\right)}\, .
\end{eqnarray} 
{\it The only requirement is that the time series is long enough so a full period of the probing signal can be observed. }

We emphasize that, in general, in an application of our method, one would have access to the absolute state of the measured agents,
\begin{eqnarray}
 x_j(t) = \delta_j^i(t) + x_j^*\, ,
\end{eqnarray}
and not directly to the deviation $\delta_j^i(t)$. 
However, the probing signal being known, we know that the deviation term vanishes at times $t=2\pi k/\omega_0$, $k\in\mathbb{Z}$, which allows to recover $x_j^*$ and to isolate $\delta_j^i(t)$. 

\subsection{Probing amplitude and frequency}\label{ssec:probing_design}
\begin{figure}
 \centering
 \includegraphics[width=\textwidth]{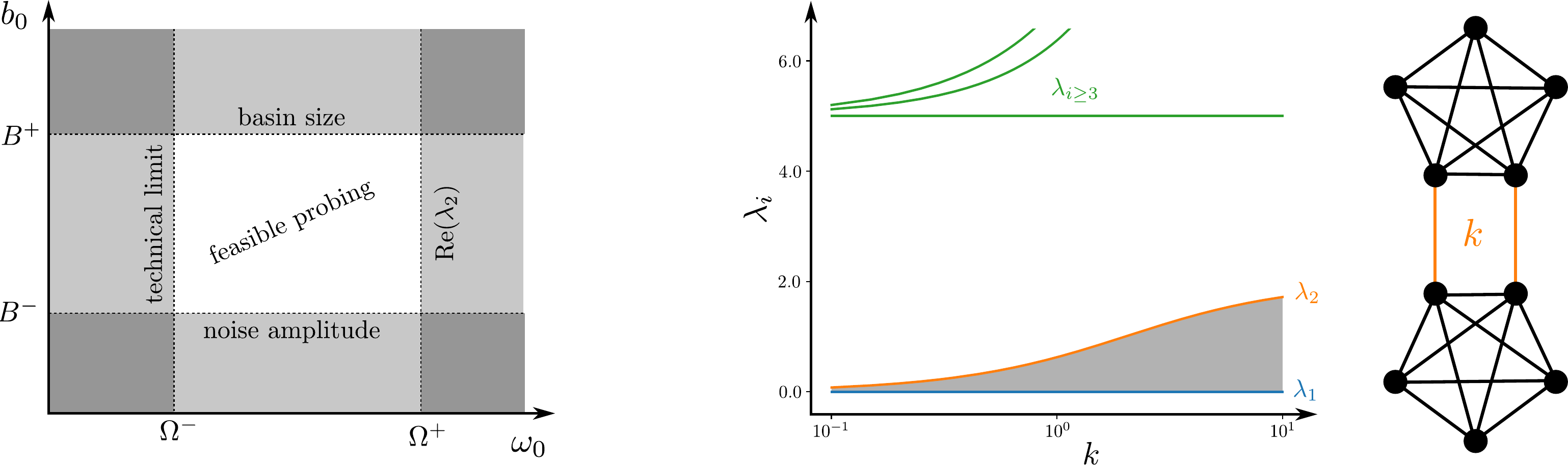}
 \caption{Left: Summary of the constraints on the probing amplitude $b_0$ and frequency $\omega_0$. 
 The frequency might be bounded from below by technical constraints and is bounded from above by the characteristic time scales of the fixed point of the system, in order to avoid transients following the probing. 
 The amplitude is bounded from below by the ambient noise amplitude so it can be distinguished from it, and from above by the size of the basin of attraction of the fixed point, to guarantee that the system remains close to its fixed point. 
 Right: Eigenvalues of the Laplacian matrix of the network depicted with respect to the coupling strength $k$ over the two orange lines. 
 Other couplings are set to one. 
 The shaded area shows the interval of feasible probing frequency. }
 \label{fig:scheme}
\end{figure}

The probing signal needs to be design in order to satisfy the assumptions of the previous section. 
With a sinusoidal probing, we can mostly act on amplitude and frequency. 
Even though there is no general recipe to determine these characteristics, we detail here determinant factors in their choice, which we summarize in the left panel of Fig.~\ref{fig:scheme}. 

The amplitude of the signal is constraint both from above and below. 
On one side, the signal should not destabilize the fixed point of the system. 
We then required it to be sufficiently small, i.e., $b_0<B^+$, so the system is not pushed out of the basin of attraction of its steady state~\cite{Pai81,Chi89}. 
Determining such upper bound on the probing amplitude requires a knowledge of the system at hand and cannot be done in full generality. 
We refer to~\cite{Men13,Sch17,Del17b} for a discussion of the notion of \emph{basin stability}, that is relevant here.

On the other side, if the system is subject to noise, then the amplitude of the probing needs to be sufficiently large so it can be distinguished from the noise, i.e., $b_0>B^-$. 
One way to determine the required amplitude is to analyze the noise to which the system is subjected, by taking measurements (which is possible at least at one agent by assumption) before injecting the probing signal. 

Note that under our assumptions, it would be unrealistic to have $B^-$ close to or larger than $B^+$, otherwise the noise would drive the system close to the boundary of the basin of attraction and eventually destabilize it. 
A system that spontaneously jumps from one basin of attraction to another cannot be considered as in steady state and is then outside the scope of this manuscript. 
There is then room to choose an appropriate probing amplitude. 

For the probing frequency, even though there is no theoretical lower limit, any implementation of our method will have some technical limitations. 
First, these limitations depend on the device or process that is implemented to inject the probing signal into the system, which might not be able to generate a signal with arbitrarily low frequency. 
Second, the steady state of a system might change over long time scales (compared to the system's response to disturbances). 
For instance, power grids can be considered at steady state at the time scale of a few minutes, but their operating state might significantly change over an hour. 
In such a case, if the steady state of the system changes within one period of the probing, our method could lead to spurious estimates. 
In general, there is then a lower bound $\Omega^->0$ to the probing frequency. 

The upper limit to the probing frequency $\Omega^+$ depends, as mentioned above, on the spectrum of the Jacobian matrix $\mathbb{J}$. 
In the linear regime about a steady state, the transient following a disturbance can be decomposed on the eigenmodes of the Jacobian. 
Each component of the transient will then decay exponentially as $e^\lambda t$, where $\lambda$ is the real part of the corresponding eigenvalue of $\mathbb{J}$, which is negative if the fixed point is stable. 
Then, introducing a disturbance continuously and at a rate smaller than the eigenvalues of $\mathbb{J}$ prevents the occurence of transient, as those are decaying faster than the disturbance is introduced. 
The impact of a slow probing signal is then to continuously displace the fixed point, without generating intractable transients. 
Determining the spectrum of the Jacobian is not a trivial task in general, in particular in our framework, where we assume that we have not many information on the system. 
However, field expert are often aware of the typical time scale of their system of interest, e.g., electrical engineers know the typical time scales in the power grid, and biochemists are aware of the time needed for their reaction to reach a steady state. 
The appropriate probing frequency should then be determined by more specialized experts. 

In contrast with the bounds on the amplitude, we cannot guarantee that $\Omega^-<\Omega^+$. 
For instance, if a system is composed of two components with strong intra-connections, but weak inter-connections (e.g., a single weak link), then the smallest nonvanishing eigenvalue $\lambda_2$, also called \emph{algebraic connectivity} or \emph{Fiedler eigenvalue}~\cite{Fie73}, can be extremely close to zero, as illustrated in the right panel of Fig.~\ref{fig:scheme}. 
In such situation, it is possible that $\Omega^->\Omega^+$, which renders our method inapplicable to the whole network. 
This observation shows that our method is restricted to networked systems that are sufficiently connected (in the sense of the algebraic connectivity), or to sufficiently connected subnetworks if weak links are present. 

{\bf Remark.}
{\it If one considers system with higher-order dynamics in Eq.~(\ref{eq:dyn}), the method still holds. 
The only difference lies in the conditions for the probing frequency to be considered as "small", in which case, one can verify that higher-order time-derivatives do not influence the response of the system to the signal. 
A more detailed discussion about the response of second-order dynamical systems to slow disturbances can be found in~\cite{Tyl21}}

\begin{figure*}
 \centering
 \includegraphics[width=.9\textwidth]{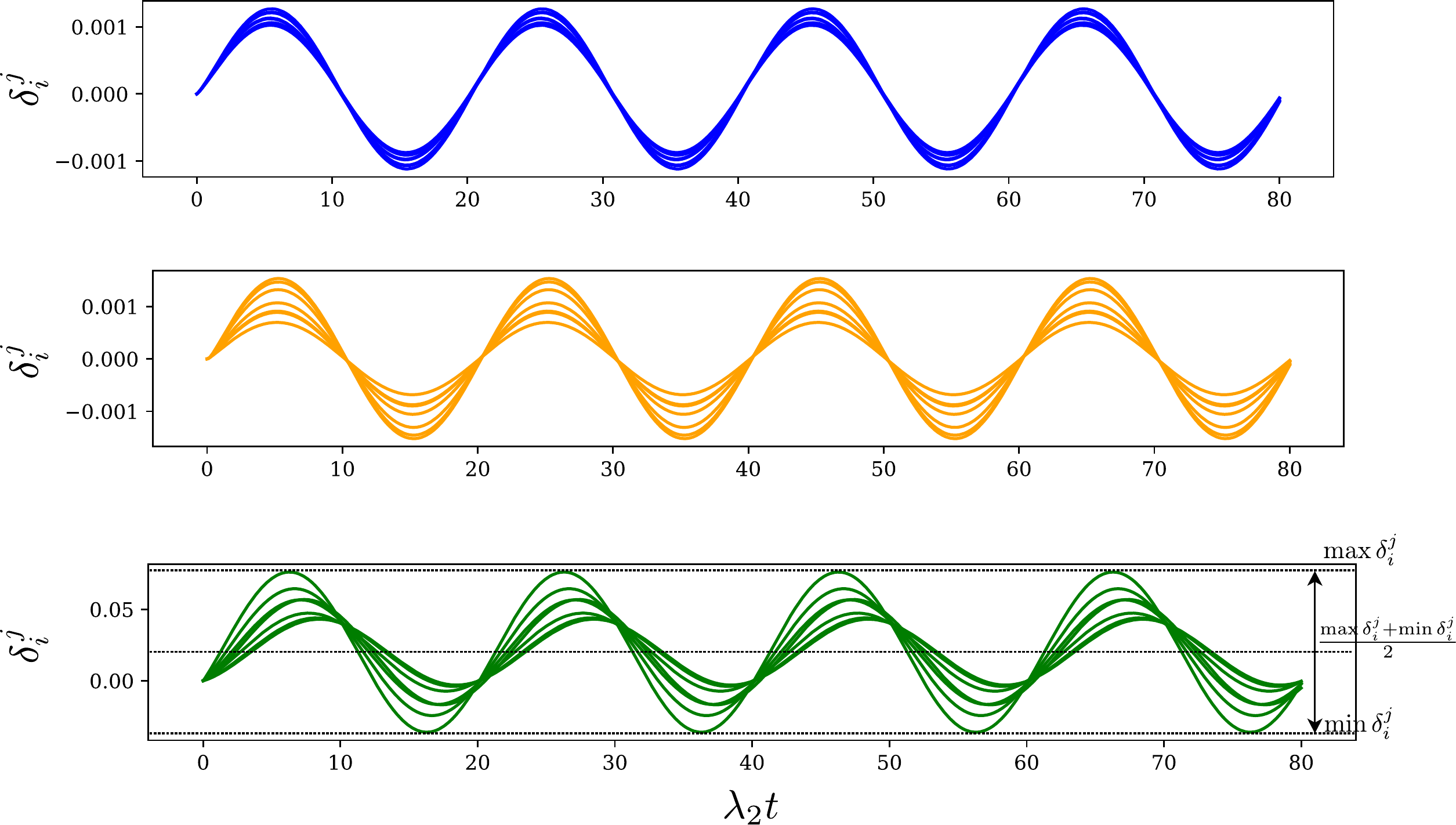}
 \caption{Trajectories obtained by probing and measuring a single node, and used to estimate the number of agent following Eq.~(\ref{eq:trajs}...)
 We consider an Erd\H{o}s-R\'enyi graph (first row) with $m=21444$ edges, a Small-World network (second row) with $m=38090$ edges, and rewiring probability $p=0.01$, and the network of the PanTaGruEl model of the European electrical grid (thrid row) with $m=4955$ edges~\cite{Tyl19,Pag19b}. 
 All networks have $n=3809$ vertices. 
 We consider a probing frequency satisfying $\lambda_2/\omega_0=20/2\pi$\,, for each network.
  The estimate of the number of agents in the system is given by the mean value around which trajectories oscillate. To get that mean value, we take the average of the maximum and minimum for each trajectories, as depicted in the third row panel.}
 \label{fig:numerics}
\end{figure*}

\section{Numerical validation}
In order to get numerical confirmation of our results, we will apply them to the Kuramoto model [$f_{ij}(x)=\sin(x)$] on three different interaction graphs. 
Each system has the same number of units, $n=3809$, but significantly different network structure: 
\begin{description}
 \item[ER] An Erd\H{o}s-R\'enyi graph with edge probability $p=0.003$ ($m=21444$ in our realization); 
 \item[SW] A Small-World constructed following the Watts-Strogatz process~\cite{Wat98}, with $m=38090$ edges and rewiring probability $p=0.01$; 
 \item[EU] The more realistic network of the PanTaGruEl model of the European high voltage grid~\cite{Tyl19,Pag19b}, composed of $m=4955$ edges. 
\end{description}
\begin{center}
\begin{table}
\begin{tabular}{ c| c c c c c c c }
$i_0$& 1& 2& 3&4 &5 &6 &7 \\ \hline 
 SW & 3807.04 & 3807.98 & 3807.07 & 3800.93 & 3806.53 & 3808.44 & 3808.11 \\  
 ER & 3808.66 & 3809.02 & 3809.02 & 3808.66 & 3808.66 & 3809.02 & 3808.66 \\
 EU & 3808.21 & 3805.52 & 3808.59 & 3808.81 & 3804.25 & 3804.22 & 3797.07
\end{tabular}
 \caption{Estimated number of agents obtained from 7 different simulations where a single was both probed and measured. 
 The corresponding trajectories are shown in Fig.~\ref{fig:numerics}. 
 All relative errors are below 0.5\%.} 
 \label{tab1}
\end{table}
\end{center}

For probing, we use a sinusoidal signal with controlled amplitude and frequency, similarly as what is typically used to identify eigenmodes in electrical networks~\cite{Pie10}. 
Figure~\ref{fig:numerics} shows the trajectories $\delta_{i}^{j}$ obtained by both probing and measuring single nodes, i.e., $i=j=i_0$\,. 
For each network (each panel of Fig.~\ref{fig:numerics}), we repeat the simulation changing probed and measured node $i_0$ over 7 different location chosen randomly. 
The estimations of the system size from these trajectories are given in Tab.~\ref{tab1}. 
One sees that our method accurately estimates the number of agent in the system as for SW and ER networks, errors of less than 8 over 3809 nodes (less than $0.21\%$) are observed and for EU network 11 nodes (less than $0.31\%$). 

We further test the robustness of our method in case of noisy conditions that reads 
\begin{eqnarray}\label{eq:dyn_lin_noise}
 \dot{\bm{ \delta}} = -\mathbb{J}(\bm{x}^*)\bm{\delta} + \bm{b} + \bm{\eta}\, ,
\end{eqnarray}
where $\bm{\eta}$ is some Gaussian white-noise acting at each node independently. 
More precisely, its moments are given by,
\begin{eqnarray}
\langle \eta_i(t) \rangle = 0 \,, \qquad \langle \eta_i(t)\eta_j(t')\rangle = \delta_{ij}\,\eta_0^2\, \delta(t-t')\,,
\end{eqnarray}
where, here $\delta_{ij}$ is the Kronecker symbol.
\begin{table}
\centering
\begin{tabular}{ c| c c c c c c c }
$i_0$& 1& 2& 3&4 &5 &6 &7 \\ \hline 
 SW & 3765.04 & 3754.33 & 3780.57 & 3733.48 & 3880.88 & 3781.9 & 3770.02 \\
 & (1.15\%) & (1.44\%) & (0.75\%) & (1.98\%) & (1.89\%) & (0.71\%) & (1.02\%) \\
 \hline  
 ER & 3770.02 & 3726.06 & 3782.18 & 3741.5 & 3816.8 & 3819.44 & 3643.84 \\
 & (1.02\%) & (2.18\%) & (0.70\%) & (1.77\%) & (0.20\%) & (0.27\%) & (4.34\%) \\
 \hline
 EU & 3780.49 & 3820.92 & 3815.41 & 3816.86 & 3791.85 & 3791.28 & 3775.53 \\
 & (0.75\%) & (0.31\%) & (0.17\%) & (0.21\%) & (0.45\%) & (0.47\%) & (0.88\%)
\end{tabular}
 \caption{Estimated number of agents and relative error obtained from 7 different simulations where a single was both probed and measured within noisy conditions. 
 The noise amplitude is such that $\eta_0/b_0=0.01$\,. 
 The simulation parameters are the same as for Fig.~\ref{fig:numerics}, except for the ER network where we double the probing time to obtain more accurate results.} 
 \label{tab2}
\end{table}
Table~\ref{tab2} shows the estimated number of agents for the dynamics of Eq.~(\ref{eq:dyn_lin_noise}), with a noise amplitude at each node satisfying $\eta_0/b_0=0.01$\,. As expected, the noise reduces the precision of the estimation. The biggest error is of 165 over 3809 ($4.3\%$), for the ER network. A better estimation of the number of agents in noisy conditions might be obtain by averaging over many probed nodes, or over longer time series.

{\bf Remarks.}
{\it To obtain the estimations in Tab.~\ref{tab2}, we averaged $\delta_i^j$ over an integer number of period of the probing signal, in order to average out the effect of the noise. 
Indeed, if the probing frequency is extremely low, one might pretty long time series to get a accurate estimate. 
Such tradeoff has been discussed in Sec.~\ref{ssec:probing_design}.}

\section{Discussion and conclusion}\label{sec:conclusion}
For diffusively coupled systems, we improved the current state-of-the-art methods to estimate the size of a complex dynamical system, based on one time series measurement solely. 
The computation cost is low and independent of the system's size. 
On one hand, our approach relies on a series of assumptions that prevent it to be applicable in full generality, to any group of diffusively coupled agent. 
On the other hand, the simplicity of the method, its low computational cost, and its scalability render it extremely valuable in contexts where it applies. 
Future work will aim, in particular, at extending the domain of applicability of this method. 

In our opinion, the performance of our approach (in terms of cost of data acquisition and computation) is close to be as optimal as possible.  
Indeed, in order to get information about the system, one needs to measure something, i.e., at least one output, which is what we do. 
Furthermore, to be able to analyze the output of the system, the observer needs some information about the input that triggered the response. 
Identifying characteristics of the system from measurements solely can hardly rely on fewer data. 
The method would be improved by relying on shorter time series, which can be long in our case, due to the low frequency of the probing signal, but in such a case, one would need to rely on a completely different approach. 

\section*{Acknowledgments}
We thank Philippe Jacquod for useful and stimulating discussions.
The authors were supported by the Swiss National Science Foundation under grants 2000020\_182050 (MT) and P400P2\_194359 (RD). 
RD acknowledges support by ETH Z\"urich funding.


\end{document}